# Sampling and Reconstruction of Sparse Signals in Shift-Invariant Spaces: Generalized Shannon's Theorem Meets Compressive Sensing

Tin Vlašić, *Student Member, IEEE,* and Damir Seršić, *Member, IEEE*

*Abstract*—This paper introduces a novel framework and corresponding methods for sampling and reconstruction of sparse signals in shift-invariant (SI) spaces. We reinterpret the random demodulator, a system that acquires sparse bandlimited signals, as a system for the acquisition of linear combinations of the samples in the SI setting with the box function as the sampling kernel. The sparsity assumption is exploited by the compressive sensing (CS) paradigm for a recovery of the SI samples from a reduced set of measurements. The SI samples are subsequently filtered by a discrete-time correction filter to reconstruct expansion coefficients of the observed signal. Furthermore, we offer a generalization of the proposed framework to other compactly supported sampling kernels that span a wider class of SI spaces. The generalized method embeds the correction filter in the CS optimization problem which directly reconstructs expansion coefficients of the signal. Both approaches recast an inherently continuous-domain inverse problem in a set of finite-dimensional CS problems in an exact way. Finally, we conduct numerical experiments on signals in polynomial B-spline spaces whose expansion coefficients are assumed to be sparse in a certain transform domain. The coefficients can be regarded as parametric models of an underlying continuous-time signal, obtained from a reduced set of measurements. Such continuous signal representations are particularly suitable for signal processing without converting them into samples.

*Index Terms*—B-spline, compressive sensing, inverse problems, sampling theory, shift-invariant spaces, sparse signal recovery

## I. Introduction

SAMPLING theorems are an essential tool that allows for processing of real-world signals on a digital processor. Due to its elegance and practicality, the Nyquist-Shannon theorem [1], [2] is the most prevalent sampling theorem. It states that a signal must be sampled at the rate that is at least twice the highest frequency contained in the signal. However, real-world signals are rarely exactly bandlimited and can often be much better represented in alternative bases [3], [4] other than Fourier. This led to a generalization of the Shannon's sampling theorem to other classes of functions. In particular, the concept extends nicely to the spline and wavelet spaces in which signals are expressed as a linear combination of the integer shifts of a generator [5]–[7]. Sampling of signals in shift-invariant (SI) spaces resembles the Shannon's sampling theorem with additional discrete-time correction filtering of samples which are not necessarily the pointwise values of the signal. Unlike the *sinc* function, there exist B-spline and wavelet scaling functions that have finite support, which makes the reconstruction formula in the SI sampling theorem feasible. In an SI sampling setting, the signal is sampled at least at the rate of innovation, and similarly to the conventional sampling theorem, it is challenging to build sampling hardware that operates at a sufficient rate when the innovation is high.

Recently, sparsity has received growing attention in the field of signal processing. It lies at the heart of compressive sensing (CS) [8]–[11], a sampling and reconstruction paradigm that has been extensively researched over the past decade. The goal of discrete CS is to recover a signal $\mathbf{x} \in \mathbb{R}^N$ from linear measurements $\mathbf{y} \in \mathbb{R}^M$ given by $\mathbf{y} = \mathbf{\Theta}\mathbf{x}$, where $\mathbf{\Theta} \in \mathbb{R}^{M \times N}$ is a sensing matrix and $M < N$. The exact recovery from such an ill-posed inverse problem is possible if the signal is $Q$-sparse, i.e., it has at most $Q \ll N$ nonzero entries, and under certain conditions [9], [12] on the sensing matrix $\mathbf{\Theta}$. While CS reduces the number of measurements sufficient for an exact recovery, and consequently the sampling rate, it increases the computational complexity of the reconstruction [13].

The vast majority of papers concerning CS focus on discrete inverse problems. Since the most of real-world signals are continuous, there are many works that extend discrete CS to the analog domain. These works mostly rely on a discretization or heuristics in order to adopt an infinite-dimensional inverse problem to a finite CS setting. The discrete model $\mathbf{\Theta}$ of a continuous measurement procedure is often an approximation, which introduces errors in the system. However, some papers are focused on solving infinite-dimensional CS problems [14]–[17]. Furthermore, hardware realizations for CS of analog signals were proposed in [18]–[22]. Alternative approaches that aim to solve continuous-domain inverse problems are in the fields of super-resolution [23]–[28] and finite-rate-of-innovation sampling [29]–[31]. These approaches are typically based on the assumption that the signal or its certain order of derivative consists of finite number of Dirac delta functions per time unit and the goal is to recover exact locations of the jumps at super-resolution.

This research was supported in part by the European Regional Development Fund under the grant KK.01.1.1.01.0009 (DATACROSS) and in part by the Croatian Science Foundation under the project IP-2019-04-6703. *(Corresponding author: Tin Vlašić.)*

The authors are with the Department of Electronic Systems and Information Processing, University of Zagreb Faculty of Electrical Engineering and Computing, Unska 3, HR-10000 Zagreb, Croatia (e-mail: tin.vlasic@fer.hr; damir.sersic@fer.hr).

This paper has supplementary downloadable material available at http://ieeexplore.ieee.org, provided by the authors. The material includes a readme file and seven audio files, which demonstrate the superiority in the sound quality of the reconstructions obtained by the proposed framework over the conventional one. This material is 21.8 MB in size.

Color versions of one or more of the figures in this paper are available online at http://ieeexplore.ieee.org.

In this paper, we propose a framework for CS of analog signals that lie in an arbitrary SI space. The inverse problem we treat is inherently continuous, as the unknown signal we want to recover is a function $f : \Omega \to \mathbb{R}$, where $\Omega \subset \mathbb{R}$ is a finite interval. Although the signal in an SI space is continuous, it is uniquely characterized by a sequence of coefficients, which makes a discretization method elegant. We show that the continuous-domain inverse problem can exactly be recast as a set of finite-dimensional CS problems. Initially, we develop a CS system based on the front-end configured as a parallel version of the random demodulator (RD) [32]. The RD's high-rate pseudo-random sequence can be reinterpreted as a combination of the box sampling functions in the conventional SI setting. Integration and sub-Nyquist sampling that follow demodulation produce samples that are linear combinations of several samples in the standard SI setting. The SI samples are recovered by solving a finite-dimensional CS problem prior to filtering by a discrete-time correction filter in order to reconstruct the signal. Furthermore, we extend the measurement procedure of the proposed framework to a wider class of SI sampling functions. The front-end configuration remains the same with an only difference in the sampling kernel. For this type of a continuous inverse problem, we propose a reconstruction procedure in which the discrete-time correction filter is embedded in the finite-dimensional CS problem. Solving the proposed CS problem directly recovers expansion coefficients that characterize the signal. We show that the discretization is exact for the SI basis functions of compact support. In numerical experiments, we demonstrate the effectiveness of the proposed framework by using the polynomial B-splines. This manuscript is a more fully developed publication based on a conference paper [33].

The main contributions of the paper are:
- We reinterpret the RD, a system that acquires sparse bandlimited signals, as a system for sampling of sparse signals that lie in a wider class of SI subspaces;
- We show that the inherently continuous-domain inverse problem can be discretized into a set of finite-dimensional problems of a CS type in an exact way by using the principles of generalized sampling in SI spaces;
- We introduce a novel framework for reconstruction of signals in SI spaces acquired by a parallel version of the RD. The signal is recovered by combining CS and the SI reconstruction procedure. The proposed framework significantly reduces the sampling rate for acquisition of sparse signals in SI spaces;
- We offer a generalization of the proposed sampling and reconstruction method to other compactly supported sampling kernels that span an SI space. In order to reconstruct acquired signals, we propose a method that embeds correction filtering in the CS optimization problem;
- We conduct the proposed framework on signals that are synthetically made $Q$-sparse and real-world signals. We provide experimental results of our methods in which the signal model is a polynomial B-spline.

The reminder of the paper is organized as follows. In Section II, we relate our work to the findings of other studies. We provide a short review on sampling in SI spaces and a formulation of a CS problem in Section III. In Section IV, we reinterpret the RD as a system that allows for CS of a wide class of signals in an SI subspace. We propose a measurement model that discretizes the continuous-domain measurements of the RD in an exact way. In Section V, a framework for reconstruction of signals in SI spaces acquired by a parallel version of the RD is proposed. In Section VI, we offer a generalization of the proposed method to other compactly supported sampling kernels. Numerical experiments are provided in Section VII and we conclude the paper in Section VIII.

TABLE I
LIST OF SYMBOLS AND NOTATIONS USED IN THE PAPER

| Symbol | Description |
| --- | --- |
| $\mathbf{x}$ | Sparse vector of coefficients |
| $\mathbf{y}$ | Vector of measurements |
| $\tilde{\mathbf{R}}$ | Correction matrix |
| $\boldsymbol{\Theta}$ | Sensing matrix |
| $\boldsymbol{\Phi}$ | Measurement matrix |
| $\boldsymbol{\Psi}$ | Sparsity matrix |
| $p_a, p_s$ | Degrees of B-spline basis functions for $a(t)$ and $s(t)$ |
| $q$ | Sparsity ratio, $q = Q/N$ |
| $H$ | Number of rows in $\tilde{\mathbf{R}}$, $H = N + 2\lceil p_s/2 \rceil$ |
| $K$ | Number of intervals in a region of interest |
| $L$ | Number of columns in $\tilde{\mathbf{R}}$, $L = N + 2\lceil p_a/2 \rceil$ |
| $M$ | Number of measurements in $\mathbf{y}$ |
| $N$ | Dimension of sparse vector $\mathbf{x}$ |
| $Q$ | Number of nonzeros in $\mathbf{x}$ |
| $T$ | Period of basis functions |
| $\rho$ | Rate of innovation of $f(t)$ |
| $\tau$ | Integration interval in random demodulator, $\tau = NT$ |
| $a(t)$ | Signal generator |
| $b^p(t)$ | B-spline of degree $p$ |
| $f(t)$ | Observed signal |
| $s(t)$ | Sampling kernel |
| $z_i(t)$ | Demodulating signal |
| $\mathcal{A}, \mathcal{S}$ | SI subspaces spanned by the shifts of $a(t)$ and $s(t)$ |
| $c[n]$ | Sequence of samples in SI sampling scheme |
| $d[m]$ | Sequence of expansion coefficients of $f(t)$ |
| $h[n]$ | Correction filter |
| $r_{sa}[n]$ | Cross-correlation sequence between $s(t)$ and $a(t)$ |
| $\delta[n]$ | Kronecker delta impulse |
| $\phi_i[n]$ | Sequence of expansion coefficients of $z_i(t)$ |

The list of important symbols and notations used in the proposed paper is summarized in Table I.

## II. RELATED WORK

The random demodulator [32], [34], [35] is a sampling system that is used to acquire sparse bandlimited signals. The RD demodulates a signal by multiplying it with a high-rate pseudo-random sequence of $\pm 1$s, referred to as the chipping sequence. The chipping sequence switches between the levels $\pm 1$ at the Nyquist rate. The demodulated signal is integrated between taking two successive samples at a low rate. The mixer that runs at the Nyquist rate can be easily designed using inverters and multiplexers [32]. The modulated wideband converter (MWC) [36] is a CS-type front-end for acquisition of multiband signals with unknown band locations. Basically, the MWC is a parallel version of the RD with lowpass filters instead of the integrators that follow demodulation. The authors consider the MWC for sampling of a sparse signal supported on $N$ frequency bands, with at most $Q \ll N$ bands active. The idea behind the RD and the MWC is aliasing of the spectrum, in such a way that each spectrum band appears in the baseband, before sampling at a low rate. The recovery is possible due to the sparse harmonic or band support.

In [37] and [38], Eldar and Mishali introduce methods for CS of analog signals in a union of SI subspaces. Sparsity is modeled by assuming that only $Q$ out of $N$ generating functions are active. Conventionally, such signals are acquired using $N$ parallel sampling filters prior to sampling at a rate $1/T$, leading to a system with sampling rate $N/T$ [39]. The authors show how to sample such signals at a system rate much lower than $N/T$ by using $M$ sampling filters, where $2Q \leq M < N$. However, the sampling rate in each channel remains $1/T$. The method can be extended to a special case of sampling of signals that lie in an SI space spanned by a single generator with a periodic sparsity pattern, i.e., out of consecutive group of $N$ coefficients, there are at most $Q$ nonzero expansion coefficients, in a given pattern. Each integer shift of the generator in a period $NT$ is modeled as a single subspace. The unknown signal is prefiltered by a set of $M$ filters whose impulse responses are weighted linear combinations of the shifts of a function biorthogonal to the generator [37], leading to a system with sampling rate $M/(NT)$, where $2Q \leq M < N$. Such sampling filters may be quite difficult to implement in hardware and are often approximated, which leads to reconstruction errors.

In [40] and [41], the authors introduce frameworks for solving continuous inverse problems that have sparse polynomial spline solutions by using the total-variation regularization. The unknown signal is assumed to be piecewise smooth and intrinsically sparse with at most $Q$ innovations. In [40], the theorem states that there exist L-spline solutions to the considered problems with sparsity $Q \leq M - N_0$, where $M$ is a number of linear measurements and $N_0$ is a spline order. Papers [42]–[45] rely on the main results of the method proposed in [40] and use grid-based discretization strategies that lead to convex optimization problems.

In this paper, we take a different approach. First of all, we use a parallel version of the RD in order to sample and reconstruct signals in a wider class of SI spaces. We discretize the continuous measurement procedure of the RD in an exact way by using the theory of generalized sampling in SI spaces, and argue that the bandlimited signal model in the conventional CS problem that uses RD measurements is just an approximation. Additionally, we propose a framework for a discretization of a continuous-domain inverse problem when the chipping sequence in an RD is replaced with a compactly supported sampling kernel that spans an SI subspace. In contrast to the cited works in which different SI spaces induce sparsity, we use SI spaces to model the underlying analog signal and discretize the continuous-domain inverse problem. The proposed framework encompasses a wider class of signals in SI spaces that may not be sparse by themselves. Similarly to the majority of papers where the pointwise values of the signal, referred to as the expansion coefficients of the *sinc* function, are assumed sparse in a transform domain, we assume that the coefficients of the SI signal model are sparse in a certain dictionary. Even though the proposed framework is not solely limited to the polynomial B-spline function spaces, throughout the paper, we put the emphasis on them since they lead to an efficient implementation and an exact discretization. Such a signal model is an alternative to the bandlimited model in the RD and its polynomial interpretation offers many practical advantages in signal processing applications.

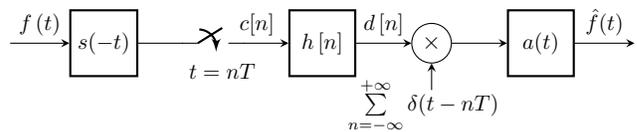

Fig. 1. Sampling and reconstruction of a signal in a shift-invariant space.

## III. BACKGROUND: GENERALIZED SAMPLING IN SI SPACES AND COMPRESSIVE SENSING

We give a short review and the most important formulations of sampling in SI spaces and CS, which we later use to successfully combine these two methods. Excellent reviews that extensively describe generalized sampling in SI spaces can be found in [5] and [7].

### A. Sampling of Signals in Shift-Invariant Spaces

Sampling in SI spaces retains the basics of the Shannon's theorem in which sampling and reconstruction are implemented by filtering. An SI subspace $\mathcal{A}$ of $L_2$ is spanned by the shifts of a generator $a(t)$ with period $T$ [5], [7]. Any signal $f(t) \in \mathcal{A}$ has the form [5]:

$$f(t) = \sum_{m \in \mathbb{Z}} d[m] a(t - mT), \quad (1)$$

where $d[m]$ are expansion coefficients that characterize the signal. Notice that the expansion coefficients $d[m]$ are not necessarily pointwise values of the signal. Shannon's theorem is a special case of SI sampling in which the generator $a(t)$ corresponds to the *sinc* function. However, signals are often better represented in other SI spaces such as splines [4], [46], [47] and wavelet scaling functions [3].

In order to guarantee a stable SI sampling theorem and a unique signal representation, $a(t)$ is typically chosen to form a Riesz basis [7]. A set of functions $\{a(t - mT)\}$ generate a Riesz basis if it is complete and there exist two positive constants $\alpha > 0$ and $\beta < \infty$ such that [5]

$$\alpha \|\mathbf{d}\|_{\ell_2}^2 \leq \left\| \sum_{m \in \mathbb{Z}} d[m] a(t - mT) \right\|_{L_2}^2 \leq \beta \|\mathbf{d}\|_{\ell_2}^2, \quad (2)$$

where $\|\mathbf{d}\|_{\ell_2}^2 = \sum_m |d[m]|^2$ is the squared $\ell_2$-norm of the coefficients $d[m]$. Riesz bases provide linear independence of the basis functions and a property that a small modification of $d[m]$ results in a small distortion of the signal [5], [7].

The SI sampling framework shares a similar sampling scheme (see Fig. 1) with the Nyquist-Shannon theorem. By analogy with antialiasing, in the SI sampling scheme the unknown signal is prefiltered by a sampling filter $s(-t)$ [5], [7]. The shifts of the sampling kernel $s(t)$ form a Riesz basis and span an SI subspace $\mathcal{S}$. Uniform sampling of rate $1/T$ follows the prefiltering stage. The sampling rate is determined by the number of degrees of freedom in the signal or the rate of innovation [29], which is in an SI sampling setting dictated by the number of expansion coefficients $d[m]$ per time unit that uniquely characterize the signal. Samples $c[n]$ of the signal $f(t)$ defined in (1) are given by:

$$c[n] = \int_{-\infty}^{\infty} f(t)s(t-nT)dt \triangleq \langle f(t), s(t-nT) \rangle, \quad (3)$$

where $\langle \cdot, \cdot \rangle$ is the conventional $L_2$-inner product. The SI samples $c[n]$ in (3) are equal to

$$\begin{aligned} c[n] &= \left\langle \sum_{m \in \mathbb{Z}} d[m] a(t-mT), s(t-nT) \right\rangle \\ &= \sum_{m \in \mathbb{Z}} d[m] \langle a(t-mT), s(t-nT) \rangle \quad (4) \\ &= \sum_{m \in \mathbb{Z}} d[m] r_{sa}[n-m], \end{aligned}$$

where $r_{sa}[n]$ is a sampled cross-correlation sequence between the sampling kernel and the generator $\langle a(t), s(t-nT) \rangle$.

To be able to reconstruct $f(t)$ from the samples $c[n]$, the discrete-time Fourier transform (DTFT) of the sampled cross-correlation sequence, denoted by $\varphi_{SA}(e^{j\omega})$, has to satisfy a mild requirement [7]:

$$\left| \varphi_{SA}(e^{j\omega}) \right| > \alpha, \quad (5)$$

for some constant $\alpha > 0$. The sequence of samples $c[n]$ in (4) have the DTFT given by $C(e^{j\omega}) = D(e^{j\omega})\varphi_{SA}(e^{j\omega})$, where $C(e^{j\omega})$ and $D(e^{j\omega})$ denote the DTFT of $c[n]$ and $d[m]$, respectively. Consequently, a recovered sequence $\hat{d}[n]$ of the expansion coefficients is obtained by discrete-time filtering of $c[n]$ with a correction filter $h[n]$ determined by [5], [7]

$$H(e^{j\omega}) = \frac{1}{\varphi_{SA}(e^{j\omega})}. \quad (6)$$

Notice that $r_{sa}[n]$ is the Kronecker delta impulse $\delta[n]$ in the case of orthogonal and biorthogonal functions $a(t)$ and $s(t)$, thus $H(e^{j\omega}) = 1$. For example, $a(t)$ and $s(t)$ are orthogonal in the Shannon's theorem. For compactly supported B-spline functions $s(t)$ and $a(t)$, $r_{sa}[n]$ has only a few nonzero entries around $n = 0$, which leads to efficient realizations of the reconstruction procedure [5]. Finally, a reconstruction $\hat{f}(t)$ of $f(t)$ is obtained by modulation of the recovered coefficients $\hat{d}[n]$ with an impulse train with period $T$, followed by filtering with a corresponding analog filter $a(t)$.

### B. Compressive Sensing

Basically, an inverse problem is to recover a signal $f$ from a finite set of noisy measurements $\mathbf{y} = (\mathbf{z}(f) + \mathbf{e}) \in \mathbb{R}^M$, where $\mathbf{z}(f) = [\langle z_1, f \rangle \ldots \langle z_M, f \rangle]^T$ are noise-free linear measurements and $\mathbf{e}$ is an additive noise term that is usually assumed to be independent of the signal. Most real-world signals are continuous and the number of measurements is finite, thus the inverse problem is ill-posed. The conventional approach to solve such an inverse problem is to select some finite-dimensional reconstruction space $\mathcal{R}$ spanned by $\{\psi_n\}_{n=1}^N$. By assuming that $f \in \mathcal{R}$ and denoting the expansion coefficients of $f$ in the basis $\{\psi_n\}_{n=1}^N$ by $\mathbf{x} \in \mathbb{R}^N$, the original inverse problem can be converted to the discretized version $\mathbf{y} = \mathbf{\Theta}\mathbf{x} + \mathbf{e}$. Matrix $\mathbf{\Theta}$ is an $M \times N$ sensing matrix whose entries are $[\mathbf{\Theta}]_{m,n} = \langle z_m, \psi_n \rangle$.

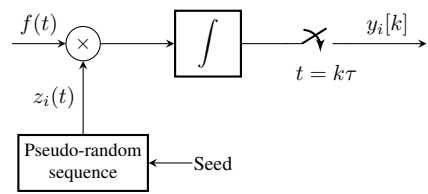

Fig. 2. Block diagram of a single channel of a bank of random demodulators.

The CS theory [8], [9] asserts that a perfect reconstruction of the signal $f$ from less than $N$ measurements is possible if $f$ is $Q$-sparse in a finite-dimensional basis $\{\psi_n\}_{n=1}^N$, i.e., $\|\mathbf{x}\|_{\ell_0} \leq Q \ll N$. The signal recovery is obtained by solving the constrained $\ell_1$-optimization problem [48]

$$\min_{\mathbf{x} \in \mathbb{R}^N} \|\mathbf{x}\|_{\ell_1} \quad \text{subject to} \quad \|\mathbf{y} - \mathbf{\Theta}\mathbf{x}\|_{\ell_2} \leq \kappa, \quad (7)$$

where $\kappa$ is a threshold parameter determined by *a priori* estimate of an error $\mathbf{e}$ such that $\kappa \geq \|\mathbf{e}\|_{\ell_2}$ [10]. The CS theory asserts that a perfect recovery of $f$ is achievable from $M$ measurements that are in the order of $Q \log(N/Q)$ under strict conditions on $\mathbf{\Theta}$, namely *restricted isometry property* (RIP) and *incoherence* [8], [9], [12], [49]. The finite-dimensional setting of CS often leads to approximations and a replacement of continuous measurements to its discrete counterparts, e.g., the continuous Fourier transform is replaced by its discrete analog. Modeling of a continuous-domain inverse problem in this way often encounters problems due to the samples discrepancy.

## IV. RD-BASED MEASUREMENT MODEL FOR SIGNALS IN SHIFT-INVARIANT SPACES

In this section, we reinterpret the RD as a system for acquisition of signals that lie in a wide class of SI subspaces. We treat the case in which the input signal is demodulated by the conventional picewise-constant signal of $\pm 1$ values. The SI function spaces are used to model the underlying continuous-domain demodulating and input signals, and the principles of generalized sampling are used to discretize the continuous measurements of an RD. The discretization method is exact and avoids approximations that lead to reconstruction errors.

We propose an acquisition system consisting of a bank of RDs as a hardware front-end. Conventionally, in the $i$-th RD channel (see Fig. 2), a bandlimited signal $f(t)$ is multiplied by a continuous-time demodulating signal $z_i(t)$ created from a pseudo-random chipping sequence of $\pm 1$. That is, $z_i(t)$ switches between the levels $\pm 1$ at least at the Nyquist rate of the signal $f(t)$. The demodulated signal is integrated prior to being sampled at a low rate $1/\tau$. The integrator is being reset after each sample $y_i[k]$ is taken.

In this paper, we consider signals that lie in a more general class of SI spaces. We assume that an input signal $f(t)$, in an SI subspace $\mathcal{A}$ of $L_2$, is spanned by a single generator $a(t)$. The signal is given by (1) with a rate of innovation $\rho = 1/T$. The signal is multiplied by a chipping sequence $z_i(t)$ that switches between levels $\pm 1$ at the rate of innovation $\rho$. Let us define the box function which is also referred to as the B-spline of degree 0:

$$b^0(t) = \begin{cases} 1, & 0 < t \leq 1 \\ 0, & \text{otherwise} \end{cases}, \quad (8)$$

which satisfies (2). In fact, the basis functions $\{b^0(t - n)\}$ are orthonormal, i.e., the constants $\alpha = \beta = 1$. If we define a sampling kernel $s(t) = b^0(t/T)$, where $1/T$ is the switching rate of $z_i(t)$, then the shifts of the sampling kernel $\{s(t-nT)\}$ span an SI subspace $\mathcal{S}$. The chipping sequence in the RD can be modeled as a signal that lie in the SI subspace $\mathcal{S}$ and is given by

$$z_i(t) = \sum_{n=0}^{N-1} \phi_i[n] \sum_{k \in \mathbb{Z}} s(t - nT - kNT), \quad (9)$$

where $\phi_i[n]$ are expansion coefficients of $\pm 1$ values. Without loss of generality, in the proposed setting we set demodulating functions to be periodic with a period $\tau = NT$ and $\phi_i[n]$ are cyclically repeated. The demodulated signal $f(t) \cdot z_i(t)$ is integrated and subsequently sampled at a low rate $1/\tau$ to obtain a sequence of measurements $\{y_i[k]\}$, for $k = 1, \ldots, K$, where $K$ is a finite number of intervals in a region of interest.

A single measurement in the $i$-th channel of the acquisition system is given by:

$$y_i[k] = \int_{k\tau}^{(k+1)\tau} f(t) z_i(t) dt, \quad (10)$$

where $k$ denotes an integration interval. By applying (1) and (9), (10) is expanded to:

$$y_i[k] = \sum_{n=0}^{N-1} \phi_i[n] \sum_{m \in \mathbb{Z}} d[m] \int_{k\tau}^{(k+1)\tau} a(t-mT) s(t-nT-k\tau) dt. \quad (11)$$

Since the integration interval covers the whole support of exactly $N$ basis functions of the SI subspace $\mathcal{S}$, the integration result in (11) for given $k$, $n$ and $m$ is equal to the sampled cross-correlation sequence

$$r_{sa}[n + kN - m] = \langle a(t - mT), s(t - nT - k\tau) \rangle. \quad (12)$$

By using (12), we can discretize the continuous-domain measurement procedure in an exact way. Thus, (11) is given by

$$y_i[k] = \sum_{n=0}^{N-1} \phi_i[n] \sum_{m \in \mathbb{Z}} d[m] r_{sa}[n + kN - m], \quad (13)$$

which, by applying (4), becomes simply

$$y_i[k] = \sum_{n=0}^{N-1} \phi_i[n] c_k[n]. \quad (14)$$

Here, $c_k[n]$ represents the conventional samples in an SI sampling setting, with the B-spline of order 0 as the sampling kernel, corresponding to the $k$-th integration interval. Using the proposed strategy, low-rate RD-based measurements $\{y_i[k]\}$ are weighted linear combinations of the SI samples $c_k[n]$.

The system's front-end consists of $M$ channels with various demodulating functions $\{z_i(t)\}_{i=1}^M \in \mathcal{S}$, where $M < N$. In each channel, the demodulated signal is integrated and sampled at a rate $1/\tau$, which leads to a system with a sampling rate $M/\tau = M/(NT)$. Not only that the proposed framework leads to an exact discretization of the RD measurement procedure, but also it allows for sampling of signals in SI subspaces with a much lower sampling rate in contrast to the conventional high-rate method described in Section III-A. The front-end of the proposed system is illustrated on the left hand side in Fig. 3. The chipping sequence in the RD scheme can be replaced by an analog filter $s(t)$ whose impulse response is a box function in this particular case. The input to the filter $s(t)$ is a modulated impulse train $\sum_n \phi_i[n]\delta(t - nT)$.

The measurement procedure of the system with additive noise is given in a matrix form by

$$\begin{bmatrix} y_1[k] \\ \vdots \\ y_M[k] \end{bmatrix} = \begin{bmatrix} \phi_1[0] & \cdots & \phi_1[N-1] \\ \vdots & \vdots & \vdots \\ \phi_M[0] & \cdots & \phi_M[N-1] \end{bmatrix} \cdot \begin{bmatrix} c_k[0] \\ \vdots \\ c_k[N-1] \end{bmatrix} + \mathbf{e}_k$$

or

$$\mathbf{y}_k = \mathbf{\Phi} \mathbf{c}_k + \mathbf{e}_k, \quad (15)$$

where $\mathbf{\Phi}$ is an $M \times N$ measurement matrix consisting of $\pm 1$ values and $\mathbf{e}_k \in \mathbb{R}^M$ is an unknown error term. Furthermore, $\mathbf{y}_k$ and $\mathbf{c}_k$ are vectors of the measurements and SI samples, respectively. Since $M < N$, the proposed finite-dimensional inverse problem is underdetermined. In the next section, we show how to recover the SI samples from an underdetermined set of equations by using the combination of CS and the reconstruction procedure in generalized SI sampling.

In the conventional RD setting, the $\mathbf{c}_k$ samples in (15) are considered as pointwise values $f(nT + k\tau)$ of the signal, which is based on the assumption that the underlying signal is bandlimited, i.e., $a(t) = sinc(t/T)$. By considering the measurement model proposed in this paper, such an inverse problem is just an approximation since $a(t) = sinc(t/T) \not\perp s(t) = b^0(t/T)$, and thus $r_{sa}[n] \neq \delta[n]$ and $c_k[n] \neq d[n] = f(nT + k\tau)$. However, the conventional RD problem is exact if $a(t)$ is assumed to be the box function, i.e., if $a(t) = s(t) = b^0(t/T)$, since it is orthonormal to its integer shifts ($r_{sa}[n] = \delta[n]$) and since its expansion coefficients $d[n]$ correspond to the pointwise values $f(nT + k\tau)$ of the signal. The proposed framework allows one to use an SI model that suits better to the observed underlying signal and which leads to exact discretization and reconstruction.

## V. Reconstruction of Signals in SI Spaces from RD-Based Measurements

The objective is to recover SI samples $\mathbf{c}_k$ from a reduced set of linear measurements $\mathbf{y}_k$. To be able to recover the SI samples by exploiting the CS theory, the samples have to be sparse in the time or a transform domain. The vast majority of works, such as [29], [37], [40], assume that an analog signal in an SI subspace is sparse in the time domain, i.e., the signal is characterized by only a few nonzero expansion coefficients $d[m]$. Analogously, we could have assumed that the SI samples $\mathbf{c}_k$ in the proposed measurement model are sparse in the time domain for signals that are spanned by a generator with a narrow support. However, in this paper, the SI model of the underlying continuous-time signal provide an alternative to the traditional bandlimited model. We consider signals that are assumed to lie in SI subspaces, but they may not have such a low rate of innovation that would make them sparse in these subspaces. Instead, we treat the case in which the SI samples $\mathbf{c}_k$ are assumed to be sparse in a certain representation basis $\mathbf{\Psi}$. This leads to the same forward model as in the conventional RD setting, which has proven to be a reasonable assumption in many real-world CS applications.

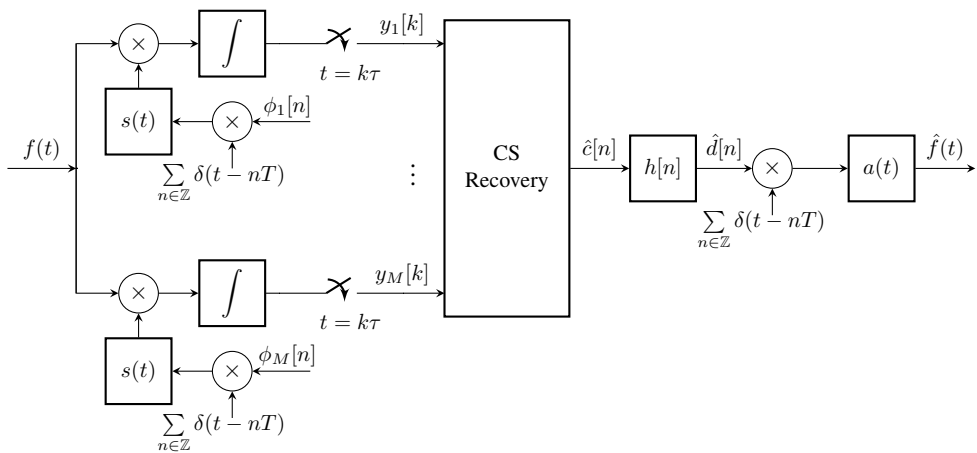

Fig. 3. Block diagram of a system for compressive sensing of signals in shift-invariant subspaces with a sampling kernel $s(t)$ and a correction filter $h[n]$.

The SI samples can be represented as $\mathbf{c}_k = \sum_{l=1}^{N} x_k[l]\psi_l$, where $\{x_k[l]\}_{l=1}^{N} = \mathbf{x}_k$ is a set of coefficients and $\psi_l$ are columns of $\mathbf{\Psi} \in \mathbb{R}^{N \times N}$. Equation (15) is then given by

$$\mathbf{y}_k = \mathbf{\Phi}\mathbf{\Psi}\mathbf{x}_k + \mathbf{e}_k = \mathbf{\Theta}\mathbf{x}_k + \mathbf{e}_k, \quad (16)$$

where $\mathbf{\Theta}$ is an $M \times N$ sensing matrix. The vector of coefficients $\mathbf{x}_k$, which is to be recovered, is exactly the same as in the conventional RD setting. The only difference is that $\mathbf{\Psi}\mathbf{x}_k$ approximates the pointwise values $f(nT + k\tau)$ of the signal in the conventional setting, and, in the proposed framework, it is equal to a set of samples $\{c_k[n]\}$ which are characteristic for generalized sampling in SI spaces with $s(t) = b^0(t/T)$.

Compressive sensing is concerned with the RIP [9], [12], [49] which, similarly to the condition in (2), secures that the sensing matrix $\mathbf{\Theta}$ preserves the geometry of the $Q$-sparse vector $\mathbf{x}$. Additionally, $\mathbf{\Theta}$ should be made of a low-coherent pair $(\mathbf{\Phi}, \mathbf{\Psi})$ of a measurement and a representation matrix to allow for a high subsampling [9]. A combination of a measurement matrix built using random entries from a certain probability distribution and any fixed representation basis has the RIP with high probability [50], [51]. Moreover, the pair of a random matrix and any fixed representation basis is largely incoherent, which makes random matrix a good choice for the measurement matrix in a CS setup [9].

Real-world signals are rarely truly sparse, but rather asymptotically sparse [11], [52]. That is, the vectors $\mathbf{x}$ of expansion coefficients have a lot of small coefficients, but only a few true zeros if any, and the signals have a structure of being far sparser at fine scales (or high frequencies) than at coarse scales.

A refined theory of CS [11], [52] offers generalized principles of incoherence and uniform random sampling, and avoids the RIP, which may be too strong an assumption in practice. The theory introduces asymptotic incoherence and sparsity, and multilevel sampling [11], [52], [53] instead of universal random subsampling. Briefly, a sensing matrix $\mathbf{\Theta}$ is asymptotically incoherent if the first few rows or columns of the matrix are large and values get asymptotically smaller as we move away from this region. Fourier/wavelet and Hadamard/wavelet transform matrices are examples of asymptotically incoherent pairs because of a high correlation of low-order frequencies and scales, and a decrease in the correlation as frequencies get higher and scales become finer. Asymptotic incoherence and sparsity structure lead to a multilevel sampling. That is, we should fully sample high coherence rows where important information about the signal is likely to be contained and as coherence starts to decrease, we can subsample gradually. The sparsity structure of the signal is exploited by using multilevel sampling of asymptotically incoherent matrices, which, in many real-world cases, leads to better reconstructions than the universal subsampling with random matrices [11], [52]. The implementation of multilevel sampling for the proposed framework is described in more detail in Section VII-B.

In this paper, we assume that the SI samples are sparse in a representation basis such as the discrete Fourier (DFT), cosine (DCT) and wavelet transform (DWT). In numerical experiments, we use the universal subsampling strategy with random matrices for synthetically $Q$-sparse signals and the multilevel sampling strategy with asymptotically incoherent matrices for real-world signals. We propose to use measurement matrices with only $\pm 1$ values in order to maintain the low complexity of the mixer design. The random Bernoulli and Walsh-Hadamard transform (WHT) matrices fit the proposed setup and their row entries are used as expansion coefficients $\phi_i[n]$ of a demodulating function $z_i(t)$.

Once the measurements are obtained, we recover the SI samples by solving the *Quadratically-Constrained Basis Pursuit* (QCBP):

$$\min_{\mathbf{x}_k \in \mathbb{R}^N} \|\mathbf{x}_k\|_{\ell_1} \quad \text{s. t.} \quad \|\mathbf{y}_k - \mathbf{\Theta}\mathbf{x}_k\|_{\ell_2} \leq \kappa, \quad (17)$$

for $k = 1, \ldots, K$, where $\kappa \geq \|\mathbf{e}_k\|_{\ell_2}$. In practice, $\mathbf{e}_k$ is often unlikely to be known. Despite that, the QCBP shows to be quite robust even when the error is underestimated [54].

A recovered set $\{\hat{\mathbf{c}}_k\}$ of the SI samples $\{\mathbf{c}_k\}$ is further used to reconstruct the signal in the conventional way for the SI sampling setting as described in Section III-A. The recovered SI samples $\{\hat{\mathbf{c}}_k\}_{k=1}^K$ in $K$ intervals of length $N$ are arranged into a sequence $\hat{c}[n]$ and filtered by the correction filter (6) to obtain expansion coefficients $\hat{d}[n]$ of the signal $\hat{f}(t) \in \mathcal{A}$. The obtained expansion coefficients $\hat{d}[n]$ are modulated by an impulse train with period $T$ and filtered by the corresponding analog filter $a(t)$. The reconstruction scheme is illustrated on the right hand side of the block diagram in Fig. 3. The proposed framework can be interpreted as a procedure of recovering of the samples from the conventional RD measurements by standard CS which is then followed by discrete-time filtering that corrects the samples in order to fit the reconstruction to a more appropriate signal subspace.

Even though, theoretically, various functions belong to the class of SI spaces, the correction filter $H(e^{j\omega})$ may be difficult to implement in practice. There are numerous filter design techniques that can be used to closely approximate the desired response of $H(e^{j\omega})$ with a finite impulse response (FIR) or an infinite impulse response (IIR) filter. However, in case when the generator $a(t)$ is a polynomial B-spline, $H(e^{j\omega})$ can be determined analytically [5]. The filter is then a non-causal IIR filter with a few coefficients corresponding to values of the cross-correlation sequence $r_{sa}[n]$. The filter can be decomposed into a causal and an anti-causal part, which leads to a forward\backward filtering [46], [47] of the recovered SI samples. Alternatively, as an impulse response of the IIR filter has a fast decay, it can be approximated by a FIR filter allowing for a continuous filtering of the SI samples with a short delay.

## VI. Generalization of the Proposed Method to Other Sampling Kernels

Physical devices often impose the sampling operator, leaving limited freedom to design the sampling strategy. That is, the sampling kernel is not always possible to be the box function proposed in Section IV. It is desirable to adopt the method proposed in the previous sections to other sampling operators which may lead to new realizations of the acquisition hardware specified for CS or may allow refined applications of CS to existing hardware. For example, one choice can be to demodulate the observed signal by a continuous-time function that has linear transition between different levels. Such a function can be modeled by the integer shifts of a B-spline kernel of order 1. Thus, we offer a generalization of the framework proposed in the previous sections, which adopts the system to other sampling kernels that form a Riesz basis for $L_2$ (2). To recover the expansion coefficients from such measurements, we embed the correction filter into the CS optimization problem in a form of a correction matrix that keeps the cross-correlation between the sampling kernel and signal generator. In general, the correction matrix can be infinite if one of the $a(t)$ or $s(t)$ kernels is of infinite support. In this section, in order to exactly recast the infinite-dimensional inverse problem into a finite CS setting, we restrict $a(t)$ and $s(t)$ to compactly supported kernels whose integer shifts form Riesz bases for $L_2$, such as polynomial B-splines, which leads to the perfect truncation of the correction matrix.

A single measurement in the $i$-th channel of the proposed acquisition system is given in (10), where $f(t)$ denotes an input signal in an SI subspace $\mathcal{A}$, $k \in \{1, \ldots, K\}$ denotes an interval of duration $\tau = NT$ and $z_i(t)$ is a demodulating signal lying in an SI subspace $\mathcal{S}$. Here, $\mathcal{S}$ is spanned by the shifts of a sampling kernel $s(t)$ from a wider class of SI spaces. Notice that the sampling kernel is not necessarily a box function $b^0(t/T)$ as it was the case in Section IV. The measurement is given by

$$y_i[k] = \sum_{n \in \mathbb{Z}} \phi_i[n] \sum_{m \in \mathbb{Z}} d[m] \int\limits_{k\tau}^{(k+1)\tau} a(t - mT)s(t - nT) dt. \quad (18)$$

Since $s(t)$ can be a sampling kernel with a wider support than $b^0(t/T)$, the basis functions may overlap each other. Thus, in general, we must include cases in which the integration interval covers more than exactly $N$ basis functions of the SI subspace $\mathcal{S}$. It follows that the integration results in (18) can not be substituted simply by the cross-correlation sequence as in Section IV, but they can be written in an infinite matrix $\mathbf{R}_k$ with rows and columns corresponding to $n$ and $m$, respectively. The continuous-domain measurement procedure is given in a discrete form by

$$y_i[k] = \sum_{n \in \mathbb{Z}} \phi_i[n] \sum_{m \in \mathbb{Z}} d[m] R_{n,m}^k, \quad (19)$$

where $R_{n,m}^k$ is an entry of $\mathbf{R}_k$ for given $n$ and $m$. Since we restricted $s(t)$ and $a(t)$ to have finite supports, the infinite $\mathbf{R}_k$ matrix is mainly filled with zeros except for a few entries around the diagonal for values of $n$ and $m$ close to $kN$. Thus, the nonzero values in $\mathbf{R}_k$ are perfectly truncated and can be concisely written in a submatrix $\tilde{\mathbf{R}}$ of finite dimensions which is universal for all integration intervals denoted by $k$. For B-splines, $\tilde{\mathbf{R}}$ is particularly easy to implement and, from now on, we assume that the sampling kernel $s(t)$ and generator $a(t)$ are polynomial B-spline basis functions. A special case occurs when both $s(t)$ and $a(t)$ are the B-splines of degree 0, then $\tilde{\mathbf{R}}$ is an $N \times N$ identity matrix. In the conventional RD-based measurement setting with $s(t) = b^0(t/T)$, the matrix $\tilde{\mathbf{R}}$ is a rectangular matrix with a cross-correlation sequence $r_{sa}[n]$ on its diagonal. Let us denote the orders of the B-spline generator and sampling kernel with $p_a$ and $p_s$, respectively. The matrix $\tilde{\mathbf{R}}$ is then of $H \times L$ size, where $H = N + 2\lceil p_s/2 \rceil$ and $L = N + 2\lceil p_a/2 \rceil$, for $p_a, p_s \in \mathbb{N}_0$. An example of the $\tilde{\mathbf{R}}$ matrix, when both $s(t)$ and $a(t)$ are polynomial B-splines of degree 1, is given in Appendix A.

By applying $\tilde{\mathbf{R}} \in \mathbb{R}^{H \times L}$, the measurement in (19) becomes

$$y_i[k] = \phi_{i,k} \tilde{\mathbf{R}} \mathbf{d}_k, \quad (20)$$

where $\phi_{i,k} \in \mathbb{R}^H$ is a row vector of the expansion coefficients corresponding to the sampling basis functions whose support is included in the $k$-th integration interval. Analogously, $\mathbf{d}_k \in \mathbb{R}^L$ is a vector of the expansion coefficients corresponding to signal basis functions whose support is included in the $k$-th integration interval. The system's front-end consists of $M < N$ channels. The demodulated signal is integrated and sampled at rate $1/\tau$, leading to a system with a sampling rate $M/\tau = M/(NT)$. The front-end of the system is illustrated on the left hand side of the block diagram in Fig. 3. Notice that contrarily to the setting in Section IV, the sampling kernel $s(t)$ of the system is not necessarily a box function $b^0(t/T)$.

The measurement procedure of the proposed system corrupted by noise $\mathbf{e}_k$ is given by

$$\mathbf{y}_k = \mathbf{\Phi}_k \tilde{\mathbf{R}} \mathbf{d}_k + \mathbf{e}_k, \qquad (21)$$

where $\mathbf{\Phi}_k$ is a measurement matrix with rows $\{\phi_{i,k}\}_{i=1}^M$. The proposed system of equations is underdetermined. To recover the expansion coefficients $\mathbf{d}_k$, we induce the sparsity and use the CS reconstruction technique. We can assume that the signal is sparse in the time domain, i.e., the vector of expansion coefficients $\mathbf{d}_k$ has only a few nonzero entries. Such an assumption is exploited in [40], [42]–[45] where the observed signals lie in the L-spline and B-spline spaces. However, in this paper, such basis functions do not provide a sparsity domain, but an SI representation model for an exact discretization of the underlying continuous-domain signal. To induce sparsity, we assume that the expansion coefficients $\mathbf{d}_k$ are sparse in a certain transform domain $\mathbf{\Psi} \in \mathbb{R}^{L \times L}$. This assumption is particularly intuitive from the wavelet theory point of view, where $\mathbf{\Psi}$ is a DWT matrix and $\mathbf{d}_k$ are projections of the signal onto the analysis scaling functions at $2^0$ resolution. The wavelet theory asserts that such coefficients are used to efficiently calculate approximation coefficients at coarser scales and wavelet coefficients of the detail signals by using the DWT [3], which has proven to yield sparse representations for wide variety of continuous-time signals. Furthermore, the $\mathbf{d}_k$ expansion coefficients can also be viewed as a sequence of numbers that is assumed to be a realization of a non-sparse Gaussian stationary process [55], for which the DCT is a universal transform which has proven to perform well in practice [56]. The expansion coefficients are represented as $\mathbf{d}_k = \mathbf{\Psi}\mathbf{x}_k$, where $\mathbf{x}_k \in \mathbb{R}^L$ is a sparse vector of coefficients in the transform domain. The problem in (21) is written as

$$\mathbf{y}_k = \mathbf{\Phi}_k \tilde{\mathbf{R}} \mathbf{\Psi} \mathbf{x}_k + \mathbf{e}_k = \mathbf{\Theta} \mathbf{x}_k + \mathbf{e}_k, \qquad (22)$$

where $\mathbf{\Theta} = \mathbf{\Phi}_k \tilde{\mathbf{R}} \mathbf{\Psi}$ is an $M \times L$ sensing matrix.

We introduced the $\tilde{\mathbf{R}}$ matrix into the inverse problem, so the recovery ability of the forward model $\mathbf{\Theta} = \mathbf{\Phi}_k \tilde{\mathbf{R}} \mathbf{\Psi}$ should be addressed. Traditionally, random matrices $\mathbf{\Phi}$ are largely incoherent with any sparsity-inducing matrix $\mathbf{\Psi}$ [9] and provide the RIP with high probability [12], [50], leading to a universal choice for measurement matrices. In the proposed framework, since $\tilde{\mathbf{R}}$ is based on the measurement setup and the choice of the underlying SI basis function, the measurement matrix is given by linking $\mathbf{\Phi}_k$ and $\tilde{\mathbf{R}}$. The $\tilde{\mathbf{R}}$ matrix is a convolution matrix consisting mainly of the cross-correlation sequence $r_{sa}[n]$ on its diagonal, corresponding to FIR filtering, and consequently affects the CS properties of the forward model in comparison to the standard $(\mathbf{\Phi}, \mathbf{\Psi})$-pair of a random measurement matrix and a sparsity matrix.

In Section VII-A, by conducting extensive experiments, we evaluate the impact of $\tilde{\mathbf{R}}$ on the perfect recovery rates of $Q$-sparse signals for various B-spline basis functions in comparison to the conventional $(\mathbf{\Phi}, \mathbf{\Psi})$-pair. We show that for many settings of the proposed framework, the $\mathbf{\Phi}_k$ matrix of random $\pm 1$s is an adequate hardware-friendly choice, which leads to a perfect recovery of $Q$-sparse signals for a comparable number of measurements as for the conventional $(\mathbf{\Phi}, \mathbf{\Psi})$-pair.

For the theoretical completeness, we offer a way to annihilate the impact of $\tilde{\mathbf{R}}$ on the CS properties of the forward model. We can introduce $\mathbf{H}$ in between of $\mathbf{\Phi}_k$ and $\tilde{\mathbf{R}}$ in (22), where $\mathbf{H}$ is the inverse IIR filter of $\tilde{\mathbf{R}}$. This corresponds to filtering of the coefficients $\phi_i[n]$ that construct $\mathbf{\Phi}_k$ by the IIR filter given in (6). Thus, $\mathbf{H}\tilde{\mathbf{R}}$ is a matrix consisting of ones on the main diagonal, except on the edges due to the interval's boundary conditions. Obviously, for such filtered random coefficients $\phi_i[n]$, the system fits traditional CS by satisfying the RIP with high probability, but the implementation of $\mathbf{\Phi}_k \mathbf{H}$ is far less convenient in real-world hardware. Fortunately, experiments show that such a procedure can be avoided in practice.

For acquisition of real-world signals, we use deterministic measurement matrices, e.g., the WHT, and the multilevel subsampling strategy [11], [52] to exploit the asymptotic incoherence and sparsity. The experiments in Section VII-B show that $\tilde{\mathbf{R}}$ does not severely affect CS properties of the forward model, and the results prove that the proposed framework in all settings yields better reconstruction results than the standard CS approach in the RD system.

The expansion coefficients are recovered by solving the QCBP optimization program for $K$ intervals:

$$\min_{\mathbf{x}_k \in \mathbb{R}^L} \|\mathbf{x}_k\|_{\ell_1} \quad \text{s. t.} \quad \|\mathbf{y}_k - \mathbf{\Theta}\mathbf{x}_k\|_{\ell_2} \leq \kappa, \qquad (23)$$

where $\kappa \geq \|\mathbf{e}_k\|_{\ell_2}$. The vector of coefficients $\mathbf{x}_k \in \mathbb{R}^L$, where $L \geq N$, is used to obtain the expansion coefficients $\hat{\mathbf{d}}_k \in \mathbb{R}^L$. The number of recovered coefficients is equal to the number of coefficients $N$ corresponding to the $k$-th integration interval plus the $L - N$ coefficients that correspond to the neighboring intervals. Next, we concatenate the expansion coefficients $\{\hat{\mathbf{d}}_k\}_{k=1}^K$ obtained in $K$ intervals of length $\tau = NT$. We apply the weighted average to the expansion coefficients on the boundaries of the integration intervals which appear in two consecutive vectors $\hat{\mathbf{d}}_k$. The weights are determined by the amount of area under a basis function in the observed integration interval. An example of the concatenation for expansion coefficients of B-splines of order 1 is given in Appendix B. The concatenated expansion coefficients $\hat{d}[n]$ are modulated by an impulse train with period $T$ and filtered by the corresponding analog filter $a(t)$ in order to reconstruct the signal $f(t)$. The reconstruction scheme is similar to the one in Fig. 3, but without a correction filter $h[n]$, which is here embedded in the CS reconstruction procedure.

## VII. Numerical Experiments

We validate the proposed system based on simulations conducted on signals that are synthetically made $Q$-sparse and real-world signals that can be seen as asymptotically sparse. All the simulations were performed in the programming language

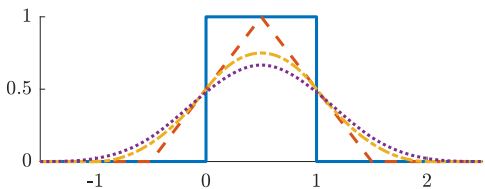

Fig. 4. B-spline basis functions of degrees $p = 0$ to 3. As $p$ increases, the basis functions flatten out and their support expands.

TABLE II
CORRECTION FILTERS FOR B-SPLINE GENERATORS OF DEGREE $p_a$

| $p_a$ | 0 | 1 | 2 | 3 |
|---|---|---|---|---|
| $H(z)$ | 1 | $\frac{8}{z+6+z^{-1}}$ | $\frac{6}{z+4+z^{-1}}$ | $\frac{384}{z^2+76z+230+76z^{-1}+z^{-2}}$ |

MATLAB and the $\ell_1$-optimization programs were solved using SPAMS v2.6 optimization toolbox [57].

### A. Q-Sparse Signals

To prove that the proposed framework perfectly reconstructs observed signals in an SI subspace under the sparsity assumption, we synthetically induced sparsity in a real-world audio signal which was previously interpolated by polynomial B-spline basis functions of various degrees.

*1) Test Signals:* A test signal is of the form

$$f(t) = \sum_{m \in O} d[m]a(t - mT), \quad (24)$$

where $a(t)$ is a generator whose shifts span a B-spline subspace $\mathcal{A}$ of degree $p_a$. Since the observed audio signal is of finite length, we restrict ourselves to a subset of $KN + 2\lceil p_a/2 \rceil$ basis functions with $m \in O$, where $O$ is a set of integers in a region of interest. For simplicity, we set the period $T = 1$. Fig. 4 illustrates the generators of the polynomial B-spline subspaces. The signal is divided in intervals of length $\tau = NT = 1024$ so that $\mathbf{d}_k \in \mathbb{R}^N$ is a vector of expansion coefficients associated with the $k$-th interval. The number of intervals is set to $K = 71$. The expansion coefficients $\{\mathbf{d}_k\}_{k=1}^K$ are synthetically made $Q$-sparse in the DCT domain. We select the sparsity ratio $q = Q/N$ to $10\%, 15\%$ and $20\%$, which correspond to $Q$: 102, 154 and 205 nonzeros, respectively.

*2) Measurements:*

*a) RD-Based Measurement:* This case simulates the measurement procedure of the conventional RD, which is described in detail in Section IV. Demodulating signals $\{z_i(t)\}_{i=1}^M$, where $M < N$, lie in the B-spline subspace of degree zero (9). Expansion coefficients $\{\phi_i[n]\}$ of $\{z_i(t)\}$ are i.i.d. random values of $\pm 1$s, which cyclically repeat with period $N = 1024$. Sampling rate in a single channel is $N$ times lower than the rate of innovation $\rho = 1/T = 1$ of the signals. Vector of measurements $\mathbf{y}_k \in \mathbb{R}^M$, that is obtained at the end of an interval $k$, is a weighted linear combination of the SI samples contained in a vector $\mathbf{c}_k \in \mathbb{R}^N$ and is given by (15).

*b) Measurement with a wider class of sampling kernels:* Here, measurements are obtained by simulating the measurement procedure of the generalized method, in which the sampling kernel $s(t)$ lies in a wider class of SI subspaces (see Section VI). In our settings, demodulating signals $\{z_i(t)\}_{i=1}^M$, where $M < N$, lie in a B-spline subspace of degree $p_s$, for $p_s = 0, 1$ or 2. The expansion coefficients $\{\phi_i[n]\}$ and the sampling rate are the same as those in the previous measurement case. The expansion coefficients $\{\phi_i[n]\}$ construct a measurement matrix $\mathbf{\Phi}_k \in \mathbb{R}^{M \times H}$, where $H \geq N$. The relation between $a(t)$ and $s(t)$ in a single integration interval is contained in $\tilde{\mathbf{R}} \in \mathbb{R}^{H \times L}$. Vector of measurements $\mathbf{y}_k \in \mathbb{R}^M$ is given by the expression in (21).

*3) Reconstruction:*

*a) CS recovery followed by a discrete-time filter $h[n]$:* We use this type of reconstruction (see Section V) exclusively for measurements acquired by the conventional RD system as described in VII-A2a. The SI samples $\mathbf{c}_k$, which are assumed sparse in the DCT domain, are recovered by solving the optimization problem (17). The threshold parameter $\kappa$ should be tuned according to the measurement error and to the type of a generator $a(t)$. Furthermore, when the generator $a(t) = s(t) = b^0(t)$, SI samples are equal to the expansion coefficients $\mathbf{d}_k$, due to the orthonormality of the basis functions. Thus, the SI samples $\mathbf{c}_k$ are ideally $Q$-sparse in the DCT domain. However, the equality $\mathbf{c}_k = \mathbf{d}_k$ does not hold for other choices of the generator $a(t)$. Consequently, the recovered SI samples $\hat{\mathbf{c}}_k$ are arranged into a sequence $\hat{c}[n]$ and are filtered by a discrete-time correction filter $h[n]$ in order to obtain the expansion coefficients $\mathbf{d}_k$, which characterize the signal in the subspace $\mathcal{A}$. Correction filters for various B-spline generators $a(t)$ are given in Table II. The correction filters are symmetric and stable, i.e., the poles are reciprocal and do not lie on the unit circle [46].

*b) CS recovery with an embedded discrete-time filter:* This reconstruction procedure is a generalization of the RD-based measurement and recovery to sampling kernels from a wider class of SI subspaces, described in detail in Section VI. The measurements are given in VII-A2b. We recover $\mathbf{d}_k$ by solving the $\ell_1$-minimization problem (23). The sensing matrix $\mathbf{\Theta}$ consists of three matrices, namely the random measurement matrix $\mathbf{\Phi}_k$, the correction matrix $\tilde{\mathbf{R}}$, and the DCT matrix $\mathbf{\Psi}$. The threshold parameter $\kappa$ in (23) is tuned according to the measurement error.

*4) Experimental Results:* We simulated every setting 20 times in order to obtain reliable results. We selected 20 sets of $\{\phi_i[n]\}_{i=1}^N$, for $n = 0, \ldots, N - 1$, that are i.i.d. random values of $\pm 1$s. These expansion coefficients of the demodulating signals, which are periodic with $NT$, were used in the simulations for all of the proposed settings. Since the test signals had been made synthetically $Q$-sparse, we were able to calculate the perfect recovery rate for various number of measurements $M$. Signal-to-noise ratio (SNR) assesses the quality of reconstructions, where the reconstruction error is treated as noise. We considered SNR of $\geq 70$ decibels to correspond to a perfect reconstruction, due to the numerical errors. In noise-corrupted measurements, the additive noise term was modeled by white Gaussian noise with variance $\sigma_n^2$. The standard deviation $\sigma_n$ was set to $5\%$ and $10\%$ of the measurement vector's standard deviation $\sigma_y$.

*a) RD-based experiments:* In this case, we used the conventional RD-based measurements and the CS recovery

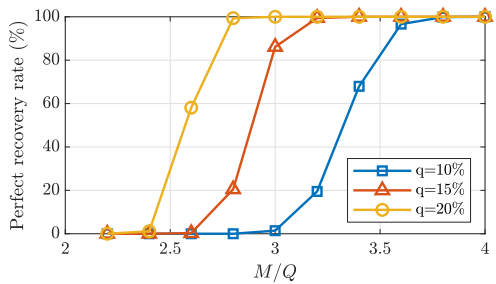

Fig. 5. **RD-based measurements and CS recovery.** Perfect recovery rates of a $Q$-sparse signal in the B-spline space of degree 0 for various number of measurements $M$ and sparsity ratios $q$.

TABLE III
RECONSTRUCTION QUALITY IN TERMS OF SNR IN DECIBELS FOR
$Q$-SPARSE SIGNALS AND RD-BASED MEASUREMENTS

|  | $q$=10% | | | $q$=15% | | | $q$=20% | | |
|---|---|---|---|---|---|---|---|---|---|
| $\sigma_n$ | 0 | $\frac{\sigma_y}{20}$ | $\frac{\sigma_y}{10}$ | 0 | $\frac{\sigma_y}{20}$ | $\frac{\sigma_y}{10}$ | 0 | $\frac{\sigma_y}{20}$ | $\frac{\sigma_y}{10}$ |
| $B_0$ | **83.93** | 17.34 | 12.36 | **75.72** | 16.61 | 12.30 | **82.17** | 16.11 | 12.36 |
| $B_1$ | 40.69 | 17.16 | 12.34 | 39.82 | 16.56 | 12.39 | 41.17 | 16.28 | 12.56 |
| $B_2$ | 37.33 | 17.34 | 12.59 | 38.39 | 16.82 | 12.66 | 37.19 | 16.53 | 12.74 |
| $B_3$ | 34.92 | **17.65** | **13.13** | 35.88 | **17.31** | **13.16** | 34.54 | **17.02** | **13.14** |

followed by a discrete-time filter $h[n]$. Since the SI samples $\mathbf{c}_k$ are ideally $Q$-sparse only for the B-spline generator $a(t)$ of degree 0, we calculated perfect recovery rates for this setting using various number of measurements $M$ and sparsity ratios $q$ of the test signal (see Fig. 5). This setting can be seen as the traditional CS problem and the simplest of the proposed settings. The graph shows that in all simulations, more than 99% of the intervals are exactly recovered for $M/Q$ ratios: 3.8, 3.2 and 2.8, which are associated to sparsity ratios 10%, 15% and 20%, respectively. These $M/Q$ ratios correspond to number of measurements $M$: 388 ($q=10\%$), 490 ($q=15\%$) and 571 ($q=20\%$). The same numbers of measurements were subsequently used in the assessment of the reconstruction results for other settings.

Reconstruction results of the proposed framework for B-spline generators $a(t)$ of degree $p=0,\ldots,3$ and various sparsity ratios are given in Table III. The reconstruction quality is the highest for noiseless settings with B-spline generator $a(t)$ of degree 0, clearly because of the sparsity property. However, even though the SI samples are not ideally $Q$-sparse in the DCT domain for B-spline generators $a(t)$ of degree 1, 2 and 3, the framework has shown to be robust in these cases, too. In noiseless experiments, reconstruction quality deteriorates as the degree of the B-spline generators $a(t)$ increases. This is mainly due to the relation of the SI samples $\mathbf{c}_k$ and the expansion coefficients $\mathbf{d}_k$, which are more similar for lower degrees of B-splines. Additionally, the correction filters progressively amplify high-frequency components as the degree of the B-spline generator increases. This results in an amplification of the reconstruction error, which is more of the high-frequency nature, and consequently deteriorates the reconstruction quality. For noise-corrupted measurements, the framework achieves similar or slightly better results than the simplest setting which is related to the traditional CS.

TABLE IV
RECONSTRUCTION QUALITY IN TERMS OF SNR IN DECIBELS FOR
$Q$-SPARSE SIGNALS ($q=15\%$) AND MEASUREMENTS WITH VARIOUS
SAMPLING KERNELS

|  | $s(t)=b^0(t)$ | | | $s(t)=b^1(t)$ | | | $s(t)=b^2(t)$ | | |
|---|---|---|---|---|---|---|---|---|---|
| $\sigma_n$ | 0 | $\frac{\sigma_y}{20}$ | $\frac{\sigma_y}{10}$ | 0 | $\frac{\sigma_y}{20}$ | $\frac{\sigma_y}{10}$ | 0 | $\frac{\sigma_y}{20}$ | $\frac{\sigma_y}{10}$ |
| $B_0$ | 75.72 | 16.61 | 12.30 | 85.52 | 19.00 | 13.82 | **84.37** | 18.95 | 13.98 |
| $B_1$ | **86.28** | **19.96** | 14.74 | **85.61** | **20.08** | **15.00** | 83.73 | **19.82** | **14.99** |
| $B_2$ | 82.31 | 19.03 | 14.59 | 48.08 | 18.46 | 14.50 | 31.60 | 17.99 | 14.34 |
| $B_3$ | 28.00 | 17.69 | 14.32 | 23.06 | 17.33 | 14.20 | 21.41 | 17.10 | 14.10 |

*b) Experiments based on the generalized method:* Here, we used measurements obtained by the generalized acquisition method and the CS recovery with an embedded discrete-time filter. First, we calculated the perfect recovery rates for settings where the B-spline generators $a(t)$ are of degree $0, 1, 2$ and 3, and sampling kernels $s(t)$ are of degree 0, 1 and 2 (see Fig. 6). The sparsity ratio $q$ is set to 15%. Notice that the perfect recovery rate of a signal in B-spline subspace of degree 0 in Fig. 6a matches the rate of a signal with sparsity ratio $q=15\%$ in Fig. 5, since these two inverse problems are equal. The same perfect recovery rate is achieved in all settings if the random values $\{\phi_i[n]\}$ are filtered with the appropriate IIR filters given in (6). However, as we see in Fig. 6, the generalized method also offers a perfect reconstruction of expansion coefficients by using random measurement matrices $\tilde{\mathbf{\Phi}}_k$ without filtering of their entries, which is a more hardware-friendly approach. It can be seen that the minimal $M$ for a perfect recovery is smaller than in the traditional problem of $(\mathbf{\Phi}, \mathbf{\Psi})$-pair for $a(t)=b^1(t)$ and $a(t)=b^2(t)$ in Fig. 6a, $a(t)=b^0(t)$ and $a(t)=b^1(t)$ in Fig. 6b and Fig. 6c. However, as the degree of B-spline basis functions $a(t)$ and $s(t)$ increases, the impact of $\tilde{\mathbf{R}}$ on the CS properties of the forward model negatively affects the minimal number of measurements $M$ that is needed for a perfect recovery.

In order to compare reconstruction results with the ones in Table III, we set sparsity ratio $q$ to 15% and the ratio $M/Q$ to 3.2. Table IV shows SNRs of the reconstructions for various degrees of the B-spline generators and sampling kernels. In noiseless experiments, while the number of measurements is large enough for a perfect recovery, the generalized method achieves much better results than the previous one. However, when the degrees of B-spline basis functions $a(t)$ and $s(t)$ increase, the reconstruction quality decreases below the reconstruction quality of the method with a correction filter. The generalized method achieves better results than the method with a correction filter in all noise-corrupted experiments. This is due to the frequency responses of the correction filter $H(z)$ and its inverse filter. While $H(z)$ amplifies reconstruction error after the CS recovery in the framework with a discrete-time correction filter, its inverse filter is lowpass and is embedded in the CS recovery of the generalized method.

*B. Real-World Signals*

Simulations for real-world signals were conducted on publicly available free samples of a high-resolution audio [58]. The original audio signal was sampled at the rate of 192 kHz.

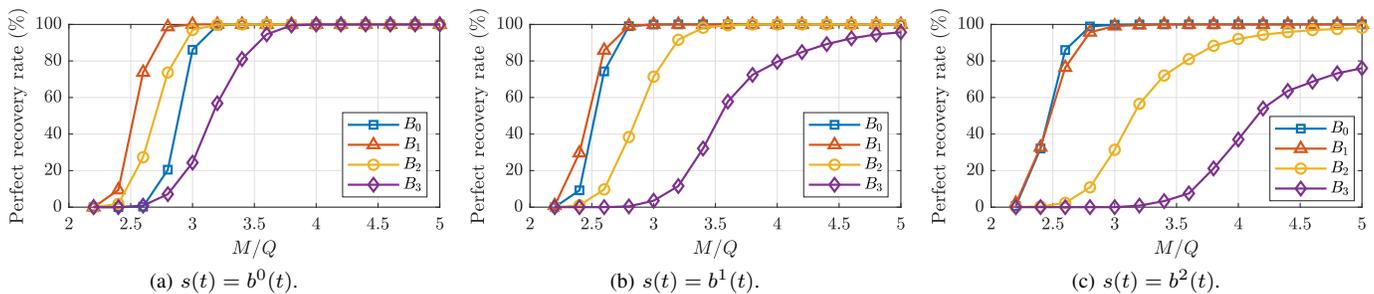

Fig. 6. **Generalized reconstruction method with a wider class of sampling kernels.** Perfect recovery rates of a $Q$-sparse signal in the B-spline space of degree $p = 0, \ldots, 3$ for various number of measurements $M$. The sampling kernel $s(t)$ is a B-spline basis function of degree: (a) 0, (b) 1 and (c) 2.

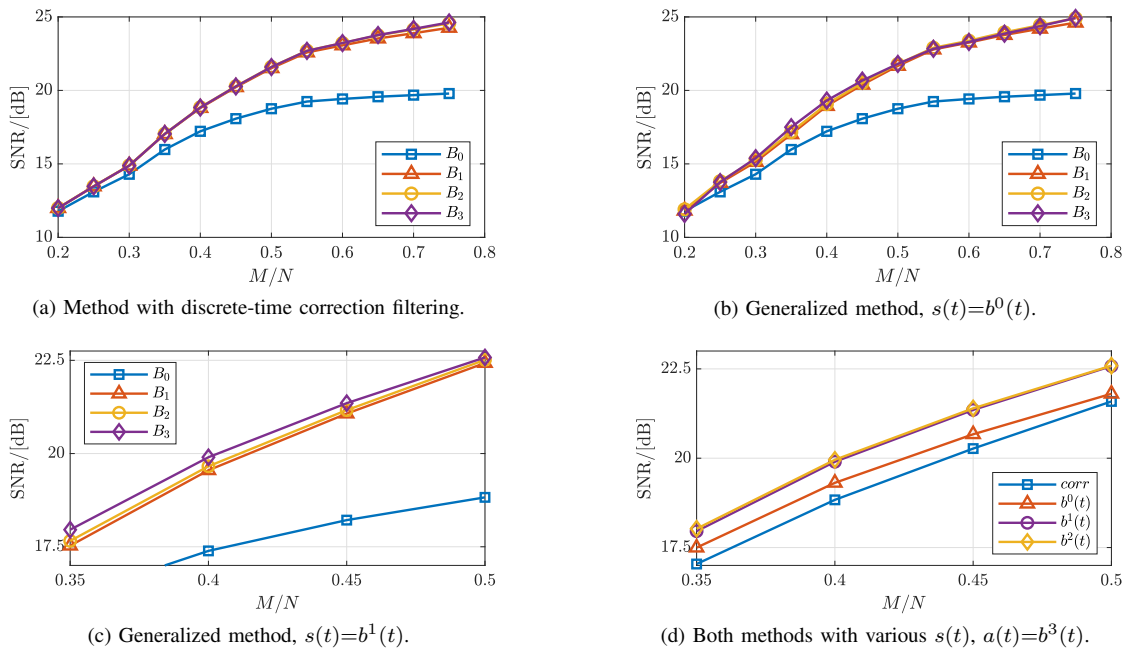

Fig. 7. **Reconstruction results of a real-world signal.** (a), (b) and (c) Reconstruction quality for various measurement settings and reconstruction methods in which the signal was assumed to lie in B-spline subspaces of degree $0, \ldots, 3$. (d) Comparison of reconstruction results of the two methods, namely the one with discrete-time correction filtering (*corr*) and the generalized method. We used $s(t)$ of degrees 0, 1 and 2, and the generator $a(t)$ was of degree 3.

The signal is asymptotically sparse in the Fourier basis with large coefficients corresponding to low-order frequencies and a lot of small coefficients, but only a few true zeros.

We simulated the measurement procedure described in Sections IV and VI with B-spline sampling kernels of order $p_s = 0, 1$ and 2. To exploit the asymptotic sparsity, we used WHT matrix rows as expansion coefficients $\{\phi_i[n]\}_{i=1}^{M}$ of demodulating functions $\{z_i(t)\}_{i=1}^{M}$, which cyclically repeat with period $\tau = NT$. The rows were picked in the multilevel-subsampling fashion [52]. We selected $\log_2(N)$ sampling levels and the number of rows in the measurement matrix assigned to the $j$-th level is $2^{j-1}$, for $j = 1, \ldots, \log_2(N)$. The amount of measurement vectors picked in each sampling level $M_j$ was uniformly assigned, i.e., $M_j = M/\log_2(N)$. Let us denote the set of indices corresponding to the $j$-th sampling level as $\mathcal{U}_j = \{U_{j-1}+1, \ldots, U_j\}$, where $0 < U_1 < \cdots < U_{\log_2(N)}$, and $N = \sum_{j=1}^{\log_2(N)} \mathcal{U}_j$. Indices corresponding to the picked rows were $I_j \subseteq \{U_{j-1}+1, \ldots, U_j\}$, $|I_j| = M_j$, and had been chosen uniformly at random. In case when $M_j > U_j - U_{j-1}$, the residual was transferred to the next sampling level.

In experimental settings, we selected the rate of innovation of demodulating signals to be 8 times lower than the sampling rate of the original audio signal, and consequently the rate of innovation of a reconstructed signal was $\rho = 1/T = 24$ kHz. The amount of coefficients corresponding to a single interval was set to $N = 1024$. Furthermore, to simulate the analog integration, we interpolated the high-resolution audio signal (between 192 kHz samples) by the zero-order hold model. Such a model is sufficient since the sampling rate of 192 kHz is a lot higher than the highest frequency of interest in audio signals and since the integration interval duration $\tau = NT$ is much longer than $T$. The small discrepancies can be modeled as the measurement noise.

The audio signal was reconstructed from a reduced set of measurements by using the two proposed methods, namely the method with discrete-time correction filtering that follows the CS recovery (see Section V) and the generalized method with an embedded correction filter (see Section VI). We simulated every setting 20 times for various indices $I_j$ and calculated the mean of the reconstruction results. The quality of reconstructions in terms of SNR are given in Fig. 7. In Fig. 7a and Fig. 7b, the graphs show that for both the reconstruction methods, representations of the signal in B-spline subspaces of degree $p_a > 0$ achieve much better reconstruction results than the representation with the B-spline of degree $p_a = 0$, which can be seen as the traditional CS method. That is, the parametric models obtained directly by the proposed methods represent signals much better than the samples obtained by the traditional method. Fig. 7c shows that as we increase the degree of a B-spline generator $a(t)$, we can accordingly expect higher SNR of a reconstruction. This is due to the support of the B-spline basis functions, which expands as the degree increases. Thus, the coefficients are recovered from additional information from neighboring intervals and the model asymptotically approaches the *sinc* reconstruction formula that fits audio signals.

The results in Fig. 7d show that the generalized method with an embedded correction filter outperforms the method with discrete-time filtering. The method with discrete-time filtering recovers exactly $N$ SI samples which are later additionally filtered with an amplifying correction filter from Table II. Contrarily, the generalized method directly recovers $L \geq N$ expansion coefficients, which is caused by the overlapping basis functions from the neighboring intervals. Thus, the latter method improves a recovery and mitigates the blocking artifacts. While the blocking artifacts are usually avoided by complex postprocessing methods, the proposed generalized method possesses the ability to reject them within the CS recovery procedure. Even though the difference between reconstruction qualities of the two methods are apparent from the SNR values, the ability to mitigate the blocking artifacts additionally enhances the reconstruction quality of the generalized method. A few reconstructions of a small segment of the audio signal are provided in the supplementary material, which clearly confirm the superiority of the generalized method over the method with discrete-time filtering and traditional CS.

Finally, the generalized method offers measurement procedures with a wider class of SI sampling kernels $s(t)$ and Fig. 7d shows that the sampling kernels of degree 1 and 2 outperform the standard RD-based measurements with $s(t)$ of degree 0.

The benefits of the proposed methods are twofold: First, the results on the real audio signal show that our discretization method is exact and leads to representations that better suit the observed analog signal. Even though the conventional RD assumes bandlimitedness, i.e., that the signal lies in an SI subspace spanned by the integer shifts of the *sinc* basis function which should perfectly fit audio signals, we show that this is an approximation which exactly corresponds to the B-spline signal model of order 0. This is clearly too coarse model for an audio signal and our framework allows to use more suitable signal spaces by elegantly linking the generalized sampling theory with the CS paradigm. Second, the recovered expansion coefficients can be regarded as a parametric model of the underlying continuous-time signal, which are in the proposed framework obtained directly from a reduced set of measurements. Such continuous signal representations offer an alternative to the traditional bandlimited model and are particularly suitable for signal processing without converting them into samples. For example, gradients and higher-order derivatives can be computed analytically, avoiding their approximation with finite difference.

## VIII. CONCLUSION

We proposed a framework for sampling and reconstruction of signals in shift-invariant spaces whose expansion coefficients are assumed to be sparse in a certain transform domain. We introduced two methods based on a combination of the principles in generalized sampling in shift-invariant spaces and compressive sensing. The methods allow for reconstruction of sparse signals with a much lower sampling rate in contrast to the traditional shift-invariant setting. The shift-invariant model of the underlying continuous-domain signal leads to the exact discretization of the inverse problem and offers an alternative and often more appropriate approach to the traditional bandlimited model. We implemented the proposed methods for sampling and reconstruction of signals that are assumed to lie in polynomial B-spline function spaces. Numerical experiments conducted on synthetically sparse data prove that our methods are robust and that a perfect reconstruction of ideally $Q$-sparse signals in B-spline function spaces is achievable for various settings. By considering the proposed measurement model, we argue that the conventional inverse problem in the random demodulator which uses bandlimited signal model is just an approximation. Reconstruction results of the real-world audio signals acquired by the random demodulator showed that the proposed framework yields higher reconstruction quality than the conventional CS setting when B-spline generators of higher orders are used.

We believe that the proposed methods can refine the way how sparse signals are acquired and modeled, as opposed to the classical bandlimited approach.

## APPENDIX A

Entries $\tilde{R}_{n,m}$ of the matrix $\tilde{\mathbf{R}}$ are determined by

$$\tilde{R}_{n,m} = \int_{k\tau}^{(k+1)\tau} a(t - mT)s(t - nT)dt. \quad (25)$$

For simplicity, we set the period of the functions to $T = 1$ and the index $k$ to 0. We choose $N = 32$ and $s(t)$ and $a(t)$ are both B-splines of degree 1. Shift-invariant subspaces $\mathcal{A}$ and $\mathcal{S}$ are identical and are spanned by $\{a(t - m)\}$ and $\{s(t - n)\}$, respectively. The SI subspace $\mathcal{A}$ is illustrated in Fig. 8. The function $a(t + 1)$ corresponds to the left integration interval ($k = -1$), but it overlaps the boundary and impacts the integration in the observed interval ($k = 0$). Analogously, the function $a(t - 32)$ overlaps the right boundary. The sampling functions $\{s(t-n)\}$ act the same way. Thus, the matrix $\tilde{\mathbf{R}}$ is an $H \times L = (N+2) \times (N+2)$ dimension matrix:

$$\tilde{\mathbf{R}} = \begin{bmatrix} \frac{1}{24} & \frac{1}{12} & 0 & 0 & 0 & \cdots & 0 \\ \frac{1}{12} & \frac{15}{24} & \frac{1}{6} & 0 & 0 & \cdots & 0 \\ 0 & \frac{1}{6} & \frac{2}{3} & \frac{1}{6} & 0 & \cdots & 0 \\ \vdots & \cdots & \ddots & \ddots & \ddots & \cdots & 0 \\ 0 & \cdots & 0 & \frac{1}{6} & \frac{2}{3} & \frac{1}{6} & 0 \\ 0 & \cdots & 0 & 0 & \frac{1}{6} & \frac{15}{24} & \frac{1}{12} \\ 0 & \cdots & 0 & 0 & 0 & \frac{1}{12} & \frac{1}{24} \end{bmatrix}. \quad (26)$$

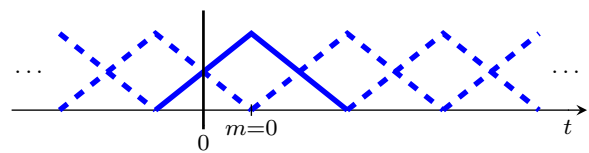

Fig. 8. Shift-invariant subspace $\mathcal{A}$ spanned by $\{a(t - m)\}$. The left boundary of the integration interval is the black line at $t = 0$. The generator $a(t)$ is illustrated by solid lines and its integer shifts by dashed lines.

Notice that the entries on the diagonal repeat when we move away from the corners and they are equal to the nonzero values of the sampled cross-correlation sequence $r_{sa}[n] = [\ldots, 0, 1/6, \underline{2/3}, 1/6, 0, \ldots]$.

## APPENDIX B

In our example, we consider the concatenation of B-spline expansion coefficients of order 1. For simplicity, we set the period of the functions to $T = 1$ and index $k$ to 0. Two basis functions, namely $a(t+1)$ and $a(t-N)$, overlap the integration interval, one at the left and another at the right boundary of the integration interval. Thus, the recovered vector of expansion coefficients is $\hat{\mathbf{d}}_0 \in \mathbb{R}^L$, where $L = N+2$. Previously, we recovered the vector of expansion coefficients $\hat{\mathbf{d}}_{-1} \in \mathbb{R}^L$ for the interval when $k = -1$. The last two entries in $\hat{\mathbf{d}}_{-1}$ and the first two entries in $\hat{\mathbf{d}}_0$ characterizes the same two expansion coefficients, namely $d[-1]$ and $d[0]$. In order to determine $d[-1]$ and $d[0]$, we apply the weighted average

$$\begin{aligned} d[-1] &= w_2 \hat{d}_{-1,L-1} + w_1 \hat{d}_{0,1} \\ d[0] &= w_1 \hat{d}_{-1,L} + w_2 \hat{d}_{0,2}, \end{aligned} \quad (27)$$

where $w_i$ are the weights. The weights are efficiently calculated from the entries of the matrix $\tilde{\mathbf{R}}$. The weights are equal to the sum of the entries in the first $L - N$ columns of $\tilde{\mathbf{R}}$. In the case when the generator $a(t)$ is the B-spline of degree 1, $L - N = 2$ and the weights are $w_1 = 1/8$ and $w_2 = 7/8$. Notice that the weights correspond to the amount of area under the basis functions in the observed integration interval and using the matrix $\tilde{\mathbf{R}}$ is an efficient way of calculating the areas.

## ACKNOWLEDGMENT

The authors would like to thank Professor Michael Wakin, Associate Editor of the IEEE Transactions on Signal Processing, and the reviewers for their valuable comments and useful suggestions that have improved the initially submitted manuscript.

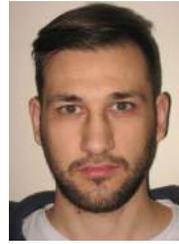

**Tin Vlašić** (S'18) was born in Zagreb, Croatia, in 1993. He received the B.Sc. and M.Sc. degrees in electrical engineering and information technology from the University of Zagreb in 2015 and 2017, respectively, where he is currently pursuing the Ph.D. degree in electrical engineering with the Faculty of Electrical Engineering and Computing.

He is currently a Swiss Government Excellence Scholar with the Department of Mathematics and Computer Science, University of Basel.

His research interests are in field of image and signal processing, specializing in particular on inverse problems, sparse modeling and sampling theory.

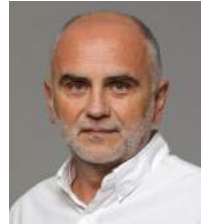

**Damir Seršić** (M'99) received the Diploma, Master, and Ph.D. degree in technical sciences, electrical engineering from the University of Zagreb, Zagreb, Croatia, in 1986, 1993, and 1999, respectively. He is promoted full professor at the University of Zagreb Faculty of Electrical Engineering and Computing.

Professor Seršić was a visiting researcher at the Technical University Vienna, Republic of Austria, in 1995, as well as at the Colorado State University, Fort Collins, Colorado, USA, in 2012.

His current research interests include theory and applications of wavelets, advanced signal and image processing, adaptive systems, blind source separation, and compressive sensing. Dr. Seršić is a member of the European Association for Signal Processing. From 2006 to 2008, he served as the Chair for the Croatian IEEE Signal Processing Chapter.